\documentclass{aa}  

\usepackage{graphicx}
\usepackage{txfonts}
\usepackage{hyperref}
\begin{document}

   \title{Two new magnetic cataclysmic variables discovered in the 3XMM catalogue}


   \author{N. A. Webb
          \inst{1}
          \and
          A. Schwope\inst{2}
          \and
          I. Zolotukhin\inst{1,3,4}
          \and
          D. Lin\inst{5}
          \and
          S. R. Rosen\inst{6}}

   \institute{IRAP, Universit\'e de Toulouse, CNRS, UPS, CNES, Toulouse, France\\
              \email{Natalie.Webb@irap.omp.eu}
\and
Leibniz-Institute for Astrophysics Potsdam (AIP), An der Sternwarte 16, 14482, Potsdam, Germany
\and
Sternberg Astronomical Institute, Moscow State University, Universitetskij pr., 13, 119992, Moscow, Russia
\and
Special Astrophysical Observatory of the Russian Academy of Sciences, Nizhnij Arkhyz 369167, Russia
\and
University of New Hampshire, 8 College Road, Durham, NH 03824-2600, U.S.A.  
\and
Dept. of Physics and Astronomy, University of Leicester, Leicester, LE1 7RH, UK             }

   \date{Received ; accepted }

 
  \abstract
   {X-ray catalogues provide a wealth of information on many source types, ranging from compact objects to galaxies, clusters of galaxies, stars, and even planets. Thanks to the huge volume of X-ray sources provided in the 3XMM catalogue, along with many source specific products, many new examples from rare classes of sources can be identified.}
   {Through visualising spectra and lightcurves from about 80 observations included in the incremental part of the 3XMM catalogue, 3XMM-DR5, as part of the quality control of the catalogue, we identified two new X-ray sources, \object{3XMM J183333.1+225136} and \object{3XMM J184916.1+652943}, that were highly variable. This work aims to investigate their nature.}
   {Through simple model fitting of the X-ray spectra and analysis of the X-ray lightcurves of \object{3XMM J183333.1+225136} and \object{3XMM J184916.1+652943}, along with complementary photometry from the {\em XMM-Newton Optical Monitor}, {\em Pan-Starrs} and the {\em Stella/WiFSIP} and {\em Large Binocular Telescope} (LBT) spectra,  we suggest that the two sources might be magnetic cataclysmic variables (CVs) of the polar type and we determine some of their properties.}
   {Both CVs have very hard spectra, showing no soft excess. They are both situated in the local neighbourhood, located within $\sim$1 kpc. \object{3XMM J183333.1+225136} has an orbital period of 2.15 hours. It shows features in the lightcurve that may be a total eclipse of the white dwarf. \object{3XMM J184916.1+652943} has an orbital period of 1.6 hours. Given that only a small sky area was searched to identify these CVs, future sensitive all sky surveys such as the {\em eROSITA} project  should be very successful at uncovering large numbers of such sources.}
   {}


   \maketitle
%

\section{Introduction}

Thanks to the high sensitivity and large field of view of many of today's telescopes, many new objects are discovered serendipitously. Many new sources have been found in the {\em XMM-Newton} catalogue from intermediate mass black holes \cite[e.g.][]{farr09,webb12} to  new magnetic cataclysmic variables (CVs), for example, eclipsing polars \citep{voge08,lin13b,rams09}. CVs are compact binaries composed of an accreting white dwarf and a main sequence or sub-giant companion star that fills its Roche lobe. Some CVs contain highly magnetic white dwarfs with strong magnetic fields \citep[$\sim$10$^{6-8}$ G;][]{rams07}. In these systems the matter becomes threaded along the magnetic field lines and is directed towards the poles of the white dwarf. This material heats up through strong adiabatic shocks and subsequently cools through plasma emission in the X-ray domain and optical cyclotron radiation. Two types of magnetic CVs are known, polars and intermediate polars. The former contain white dwarfs with very strong magnetic fields that prevent the accretion disc from forming  \citep[e.g.][]{rams07}. The strong magnetic field also tidally locks the white dwarf to the companion star so that the orbital period of the binary and the rotational period of the white dwarf are equal. Intermediate polars have slightly weaker fields so the accretion disc can form but its inner region is disrupted. The white dwarf magnetic moment in this case is insufficient to tidally lock the two stars and the white dwarf rotates at higher frequencies than the orbital frequency of the binary system \citep[e.g.][]{warn95,bern17}.

Magnetic CVs are interesting to study as the mass, radius, and composition of the white dwarf can be constrained \citep[e.g.][]{muka90}, and their accretion processes can be probed \citep[e.g.][]{rams04}.  Furthermore, short orbital period CVs are often faint in the optical domain, but can be bright in X-rays if they are magnetic. Using the high energy domain can then be helpful for identifying the CVs with the shortest orbital periods close to the period bounce, to help us understand how such objects evolve \citep[see ][for a review]{knig11a}.

Here we present two new X-ray sources from the 3XMM catalogue, for which we analyse the X-ray data, report on optical follow-up observations, discuss some of the system parameters, and propose a possible magnetic CV classification.

\section{Data reduction and analysis}

\subsection{X-ray data}

The two sources presented in this paper were discovered whilst visually screening spectra and lightcurves provided as a part of the 3XMM catalogue \citep{rose16,wats09}. The high time variability, showing evidence for periodic behaviour and hard spectra is intriguing, so we investigated the sources further.

The {\em XMM-Newton} observations were made as outlined in Table~\ref{tab:XMMdata}. Also given in the Table are the cameras (and the filters) with which the sources were detected. The source \object{3XMM J184916.1+652943}\footnote{\url{http://xmm-catalog.irap.omp.eu/source/206913206010002}} was detected with the MOS2 camera only as it fell at the position of the dead charged coupled device (CCD, number 6) in the MOS 1 array. The pn camera was operated in small window mode and the source was at 9.3\arcmin\ off axis, far outside the field of view.  \object{3XMM J183333.1+225136}\footnote{\url{http://xmm-catalog.irap.omp.eu/source/206939701010004}} also fell at a high off-axis angle of 7.4\arcmin, but as all three cameras were operated in full frame mode, it was detected with each of them. 

\begin{table*}
\caption{{\em XMM-Newton} data of the two CVs. The 3XMM source name is given and then the observation date, followed by the cameras with which the source was detected, along with the filter type. All cameras were in the full-frame mode. The total exposure time and the good time interval used for analysis are given in ks and the extraction radius is also provided.}             
\label{tab:XMMdata}      
\centering          
\begin{tabular}{l c c c c c  }     
\hline\hline       
Source & Date & Camera & Filter & Exposure (GTI) ks & Extraction radius (\arcsec)\\
\hline
3XMM J183333.1+225136 & 2012 Sep. 17 & MOS1 + MOS2 & Medium & 32 (26) & 30 \\
        & & pn &       Medium &    30 (21) & 20 \\
\hline
3XMM J184916.1+652943 & 2012 Nov. 10 & MOS2 & Thick & 12.5 (12.5) & 43.5 \\
\hline                   
\end{tabular}
\end{table*}

We reduced the raw {\em XMM-Newton} data using the
{\em XMM-Newton} {\tt Science Analysis Software} ({\tt SAS}, version 13.5) and the
latest calibration files at the time of the data reduction (CCFs, May 2014). The MOS data were
reduced using the {\tt SAS} task {\em emproc} and the {\tt SAS} task {\em barycen} was used to barycentre the data, using the coordinates of the source of interest.  The event lists were filtered with the
\#XMMEA\_EM flag, and 0--12 of the predefined patterns (single, double,
triple, and quadruple pixel events) were retained. For \object{3XMM J183333.1+225136} we identified periods of high background in the same way as described in the {\em XMM-Newton} {\tt SAS} threads\footnote{http://xmm.esac.esa.int/sas/current/documentation/threads/} and the good time interval is given in parenthesis in Table~\ref{tab:XMMdata}. The background was
low and stable throughout the \object{3XMM J184916.1+652943} observation, so we were able to use all of the data.  We also filtered in energy, using the range 0.2--12.0 keV.
The {\it pn} data were reduced using the `epproc' and zero to four of the
predefined patterns (single and double events) were retained, as these
have the best energy calibration. Again we used the task {\em barycen} to barycentre the data.  The background was treated in the same way as for the MOS data. We used the \#XMMEA\_EP filtering and the same energy range as for the MOS. 

The {\tt SAS} provides a task ({\em especget}) which allows the user to find an extraction region that optimises the source signal with respect to the background. We extracted the data using {\em especget} and the regions used are given in Table~\ref{tab:XMMdata}. The background was chosen from a source free
region close to the source.  To create the spectra we
rebinned the  data into 5eV bins as recommended in the SAS
threads.
We used the {\tt SAS} tasks `rmfgen' and `arfgen' to generate a
`redistribution matrix file' and an `ancillary response file', for
each spectrum. The data were binned to contain at least 20 counts per
bin.  The spectra were fitted using {\em Xspec} version 12.5 \citep{arna96}. Initially we tried a simple power law model to fit the spectra. For low signal to noise spectra, this type of model can be appropriate \cite[e.g.][]{home06}. We also tested thermal models, such as a thermal bremsstrahlung model, a mekal, with an abundance of one or an apec model, in association with absorption due to the interstellar medium ({\em tbabs} in {\em Xspec} with the \cite{wilm00} abundances), to see if there was any improvement to the fit.    

\begin{figure}
   \centering
   \includegraphics[width=6.5cm, angle=-90]{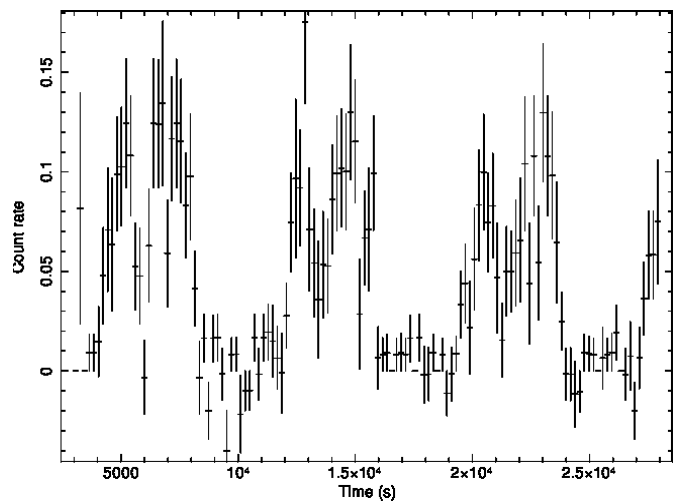}
   \caption{pn X-ray (0.2--12.0 keV) lightcurve of \object{3XMM J183333.1+225136}. The data are shown in bins of 121 s. Start time is MJD=56187.8738}
              \label{fig:CV2lc}%
    \end{figure}

\begin{figure}
   \centering
   \includegraphics[width=6.5cm, angle=-90]{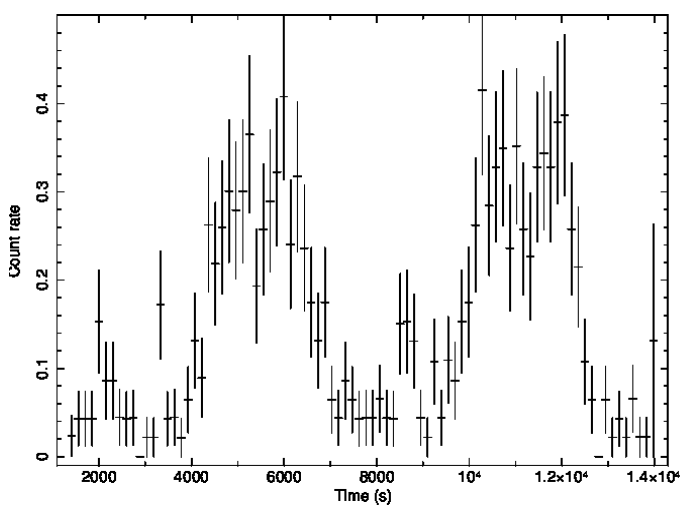}
   \caption{MOS2 X-ray (0.2--12.0 keV) lightcurve of \object{3XMM J184916.1+652943}. The data are shown in bins of  148 s. Start time is MJD=56241.7229.}
              \label{fig:CV1lc}%
    \end{figure}

The lightcurves were extracted using the same regions as for the spectra and using the shortest time bins of the camera in use, as well as using the same temporal range for the source and background lightcurves for each camera. The source lightcurves were corrected for the background using the task {\em epiclccorr}. These lightcurves can be seen in Figs.~\ref{fig:CV2lc} and \ref{fig:CV1lc}.   To search for periodicities, we applied the {\tt ftool} ‘powspec’ to the barycentred, non-background subtracted lightcurves. For \object{3XMM J183333.1+225136}, we used the pn data as it has the best time resolution. For \object{3XMM J184916.1+652943}, we used the MOS 2 data, the only one available. The period search was refined using the {\tt ftool} ’efsearch’. We then validated the periods by eye using the ftool ’efold’.  The power spectra of the two sources can be seen in Figs.~\ref{fig:1833power} and \ref{fig:1849power}.


\subsection{Optical observations}

\subsubsection{Photometric data}

The source \object{3XMM J183333.1+225136} was also observed with the {\em XMM-Newton Optical Monitor} \citep[OM,][]{maso01}. It was detected at 18$^h$33$^m$33$\fs$2 +22$^\circ$51\arcmin36$\farcs$1, coincident within the X-ray source positional errors. The observations were made in full-frame low-resolution mode (with 1$\arcsec$ binned pixels), which is 17$\times$17\arcmin\ in size. The observations were made in five different filters as detailed in Table~\ref{tab:OMdata} which presents the results taken from the pipeline processing system (PPS) data. Due to the large off-axis angle, \object{3XMM J184916.1+652943} does not fall in the OM field of view.

The positions of the two sources do not fall in the footprint of the
{\em SDSS} imaging survey \citep{alba17}. The very recent {\em Pan-STARRS} data release \citep{flew17} instead reveals probable counterparts for both objects.

Time-resolved photometry of the two sources was performed with {\em STELLA}, a
robotic 1.2m telescope equipped with WiFSIP (Wide-Field Stella Imaging
Photometer) and located at the Izana observatory in Tenerife, Spain
\citep{stra04}. An overview of the STELLA/WiFSIP photometric observations is given in 
Table~\ref{tab:optobs}. 

\begin{table}
\caption{Optical photometry with STELLA/WiFSIP}
\label{tab:optobs}
\begin{tabular}{lrrr}
BJD(TDB) & frames & T$_{\rm exp}$ (s)  & filter \\ 
\hline
\multicolumn{4}{l}{\it 3XMM J183333.1+225136}\\
2456860 &  96 &  60 & $g'$ \\
2456862 & 223 &  60 & $g'$ \\
2457173 &  85 & 180 & $g'$ \\
2457189 &  98 & 180 & $g'$ \\
\hline
\multicolumn{4}{l}{\it 3XMM J184916.1+652943}\\
2456860 & 252 &  60 & $g'$ \\
2457173 & 124 & 180 & $g'$ \\
2457189 &  50 & 180 & $g'$ \\
\end{tabular}
\end{table}

Initially, in 2014, a rather short integration of 60\,s per exposure was used,
which led to rather noisy data given the faintness of both targets. 
In 2015, an exposure time of 180\,s per frame was used. 
The latter observations, however, were affected by problems with the
cooling system of the instrument and reduced atmospheric transparency. 
These problems led to an enhanced and strongly variable dark current so that
the data quality was no better and sometimes even worse than the 2014 observations.

\subsubsection{LBT spectroscopy}

The two optical counterparts to the X-ray sources were observed with the Large Binocular Telescope (LBT),
equipped with MODS (Multi-Object Double CCD Spectrographs) on both `eyes' \citep{roth16}, in November 2016. The spectrographs were used 
in dual grating mode with a dichroic feeding blue and red spectrograph
channels at 570 nm. The gratings used for the two channels revealed
reciprocal dispersions of 0.52 \AA/pixel and 0.85\AA/pixel in the blue and the red
channels respectively, and a complete wavelength coverage between 330\,nm and
1\,$\mu$m. 
 
The data reduction was done in the standard way, using MODS-specific scripts provided by the LBT Observatory for pre-reduction (bias correction, pixel flats), and MIDAS-scripts for spectral flats, wavelength calibration, spectral extraction, and flux calibration. The estimated photometric uncertainty is about 30\% since the standard stars were not obtained on the same nights as the target stars and were taken through a different slit. The targets could only be observed at high airmass. Both targets were observed for three times 300\,s on the night of 20-21 November 2016. 

\section{Results}

\subsection{3XMM J183333.1+225136}

\subsubsection{X-ray and optical variability}

\begin{figure}
   \centering
   \includegraphics[width=10cm, angle=0]{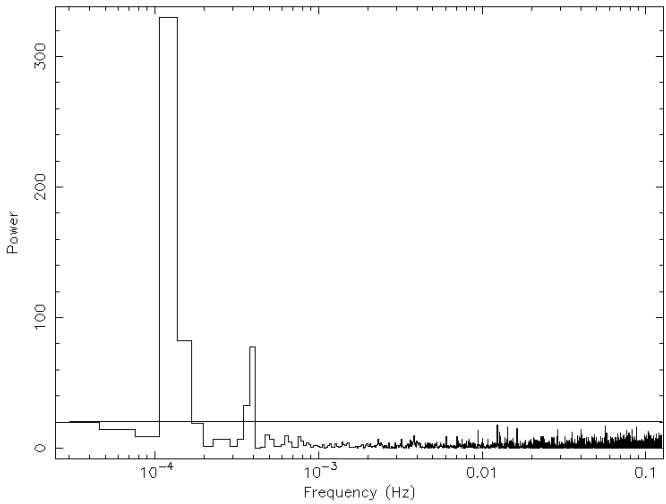}
   \caption{Power spectrum of \object{3XMM J183333.1+225136}. White noise has been subtracted. The solid horizontal line indicates a 99.95\% significance assuming pure Poisson noise.}
              \label{fig:1833power}%
    \end{figure}

\begin{figure}
   \centering
   \includegraphics[width=10cm, angle=0]{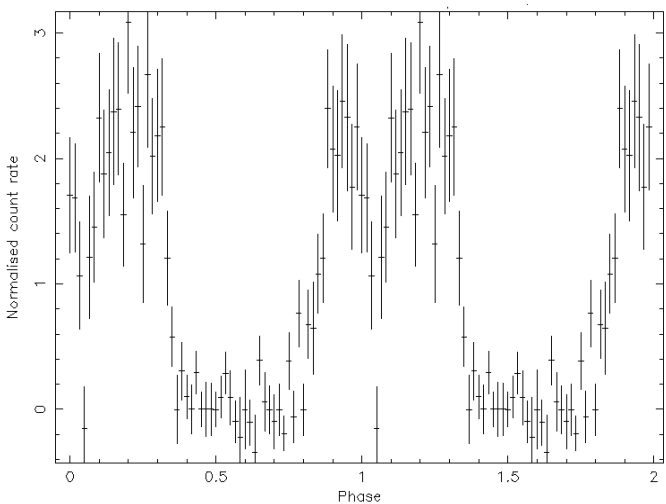}
   \caption{0.2--12.0 keV pn lightcurve of \object{3XMM J183333.1+225136}, folded on the optical period of 7715.8 s and shown with bins of 128.6 s. Two periods are shown for clarity.}
              \label{fig:CV2FoldedLC}%
    \end{figure}

Two significant peaks (above 99.95\% significance) are found in the power spectrum of \object{3XMM J183333.1+225136}. The strongest peak at P $\simeq$ 7755$\pm$30 s (3 $\sigma$ error) is likely to be the orbital period and the peak at $\sim$2585 s is P/3. We also expect to detect P/5 for a square shaped pulse profile, $\sim$1551 s, which is also detectable but at a lower significance, see Fig.~\ref{fig:1833power}.   The data appear to span three orbital periods, see Fig.~\ref{fig:CV2lc}, so we folded all the data on the X-ray orbital period, using the start time of the observations as the phase zero. We then refined the process using the optical ephemeris derived from the optical observations (see below). The folded lightcurve, shown twice for clarity, can be seen in Fig.~\ref{fig:CV2FoldedLC}.  The pulse can be seen to last for almost 0.5 of the orbit and shows a fairly steep rise lasting less than 0.1 of the orbit and an eclipse of the white dwarf, centred at $\Phi$=0.05, that is consistent with zero counts at its deepest point and appears to last for $\sim$0.09 of the orbit.  Throughout the rest of the orbit ($0.4<\Phi<0.8$) the count rate is consistent with zero (99\% confidence), when the accretion region is  hidden by the solid body of the white dwarf \citep[e.g.][]{muka17}.  

We investigated whether the lightcurve was energy dependent, as is often seen in polars \citep[see e.g.][]{maso85}.  We examined the soft (0.2--2.0 keV) and the hard (2.0--12.0 keV) X-ray bands which we found to be energy independent. The pulsed flux (f) fraction (f$_{max}$-f$_{min}$/f$_{max}$) for both bands was the same as for the combined band (100\%).

An optical counterpart was significantly detected with the OM in only the UVW1 band (see Table~\ref{tab:OMdata}), although there is some evidence for a faint counterpart in all filters, but below the detection threshold. The optical counterpart is located in a fairly crowded area of the sky where the closest source is 5\arcsec\ away. The neighbouring source was detected in the U, B, and V bands, but not in the UVW1 band.

\begin{table}
\caption{Photometric data for \object{3XMM J183333.1+225136}. The telescope, filter, exposure and (limiting) AB magnitudes are given. Upper limits are 3 $\sigma$.}             
\label{tab:OMdata}      
\centering          
\begin{tabular}{c c c c }     
\hline\hline       
Telescope & Filter & Exposure (s) & Magnitude \\
\hline
Pan-STARRS & $i'$ & 360 & 22.02$\pm0.14$ \\
Pan-STARRS & $r'$ & 120 & 22.12$\pm0.15$ \\
Pan-STARRS & $g'$ & 215 & 21.80$\pm0.10$ \\
OM & V & 2671 & $>$20.9   \\
OM & B & 3690 & $>$22.0   \\
OM & U & 3671 &$>$22.6   \\
OM & UVW1 & 4070 & 20.5$\pm$0.4 \\
OM & UVM2 & 4761 &$>$21.5   \\
\hline                   
\end{tabular}
\end{table}

The {\em Pan-STARRS} catalogue lists PSO J183333.179+225136.213 at RA(2000)=18:33:33.178, DEC(2000)=+22:51:36.16 (mean aperture magnitudes given in Table~\ref{tab:OMdata}), just 1.1\arcsec\ from the X-ray source position. Its time-resolved optical photometry confirms that it is the optical counterpart of \object{3XMM J183333.1+225136} (see below).

\begin{figure}
\resizebox{\hsize}{!}{\includegraphics[clip=]{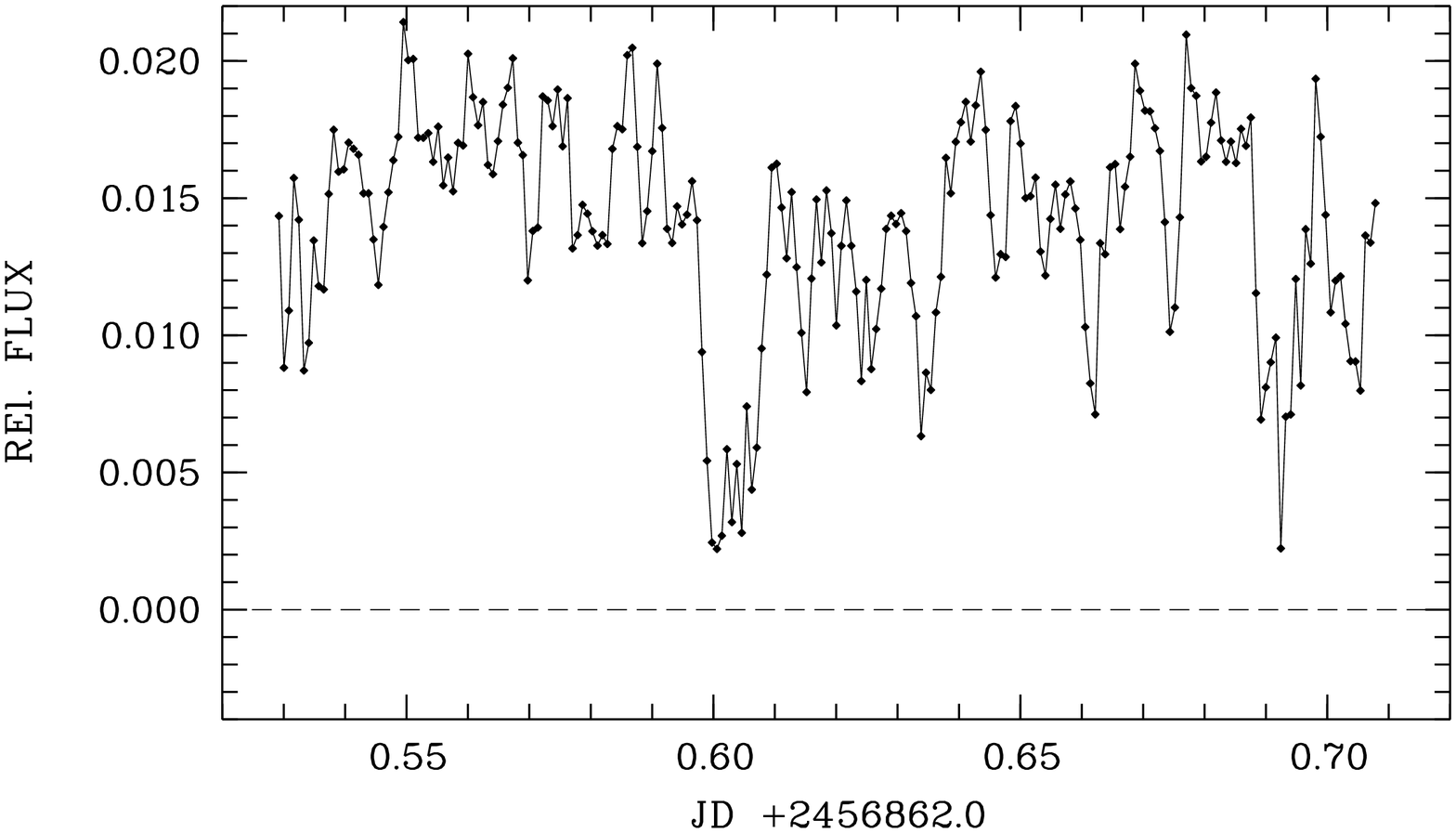}}
\resizebox{\hsize}{!}{\includegraphics[clip=]{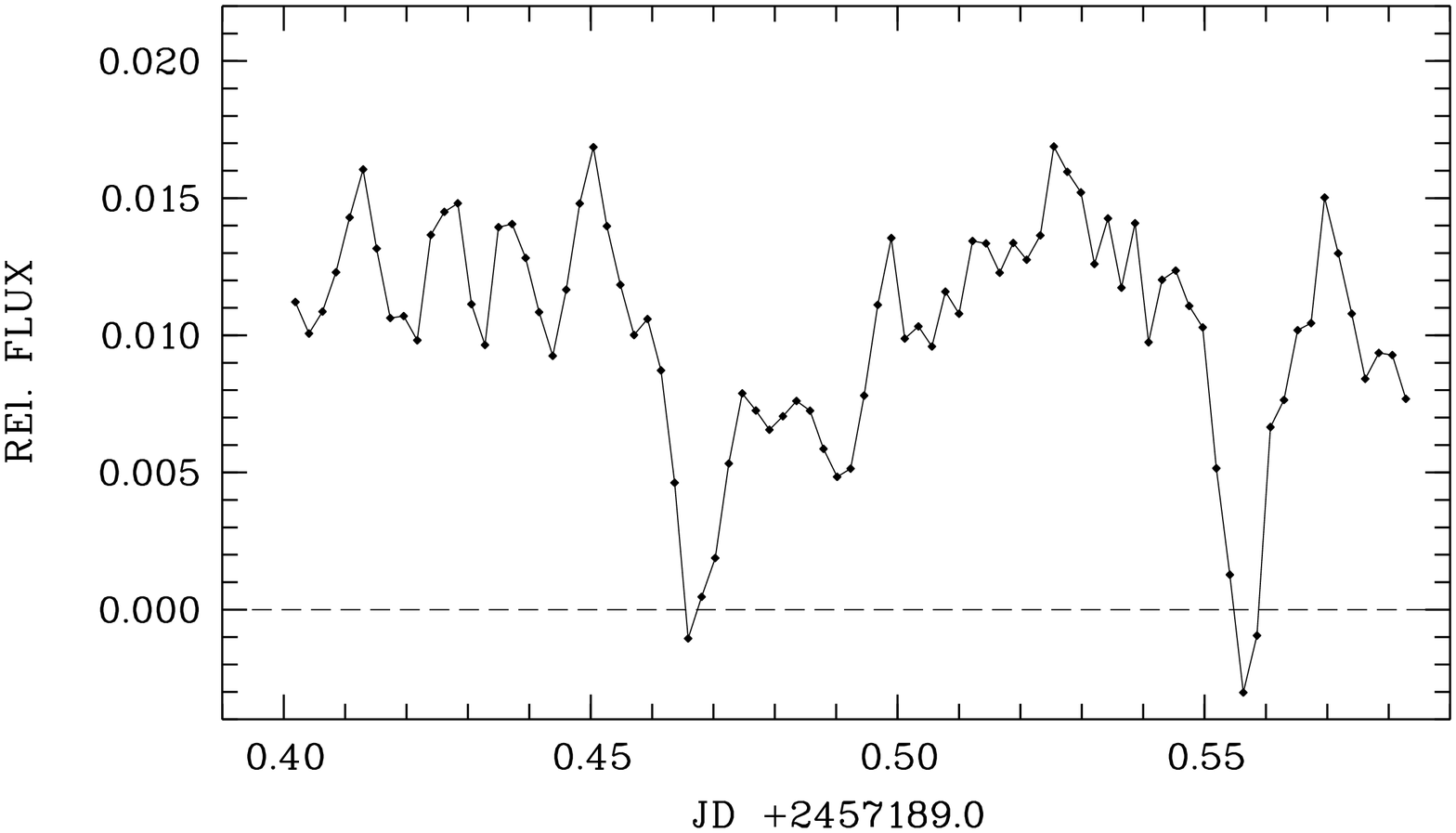}}
\caption{STELLA/WiFSIP photometry of \object{3XMM J183333.1+225136} obtained in July 2014  (top plot) and June 2015 (bottom plot). The relative flux is the flux of the optical counterpart divided by PSO\,J183332.533+225135.43, to correct for sky variations. The photometric uncertainties are of the order of the data scatter outside the eclipse.
\label{f:stella1833}}
\end{figure}

The STELLA/WiFSIP differential aperture photometry was performed with respect to the object  PSO\,J183332.533+225135.43 which has a mean aperture magnitude $g'=16.762 \pm 0.004$. This 'relative flux' corrects for sky variations. The results show a variable object, with flux changes of 100\% with a maximum magnitude of $\sim$21 mag (Fig.~\ref{f:stella1833}).  Each observing run revealed
an eclipse (or eclipse-like feature) lasting about ten minutes
(see Fig.~\ref{f:stella1833}), similar to the X-ray eclipses observed.

All lightcurves appear to indicate a pulse preceding the
eclipse. The large photometric uncertainties are of the size of the
data scatter outside the eclipse. They prevent a more detailed
characterisation of the lightcurve as far as, for example,~the length of the pulse, and the phasing of the eclipse with respect to the pulse centre are
concerned. 

\begin{table}[h]
\caption{Times of optical eclipses of \object{3XMM J183333.1+225136}. The colon after the given cycle number indicates a possible one cycle alias. Columns are 1) the number of orbital periods after cycle 0, 2) the barycentric Julian date time given for the centre of the eclipse, 3) the error in measuring the centre of the eclipse and 4) the observed period length for that cycle (O) minus the calculated value based on the ephemeris given in Eq.~\ref{eq:1833ephem} (C) and normalised by the timing measurement error.}
\label{t:oecl}
\begin{tabular}{cccc}
Cycle & BJD(TDB) & $\Delta T$& $(O-C)/\Delta T$ \\
\hline
    0& 2456860.4645& 0.0004&0.2\\
   24& 2456862.6076& 0.0004&-0.4\\
   25& 2456862.6972& 0.0006&0.2\\
 3684:& 2457189.4724& 0.0008&0.5\\
 3685:& 2457189.5609& 0.0008&-0.5\\
\end{tabular}
\end{table}

We thus concentrate on the timing of the eclipse. Times of the centre of this
event were determined from the full width at half maximum and transformed to
barycentric Julian date (Table~\ref{t:oecl}). 

A linear ephemeris of this event was built in the following way. The
observations on 2014/07/24 covered two eclipses and their separation
was 129 min. This was sufficient to combine with the measurement obtained two
nights prior to this and gave a period of $7715.8 \pm 1.8$\,s. The uncertainty
was sufficiently small to extrapolate to the data obtained in 2015, although with a
one cycle alias. The cycle numbers given in the first column of
Table~\ref{t:oecl} refer to this best fitting initial period. Linear regression
then yields a final ephemeris for the centre of the optical
eclipse of 
\begin{equation}
\label{eq:1833ephem}
T_0 (BJD(TDB)) = 2457025.058(45) + E \times 0.08930717(17)
.\end{equation}
The numbers in parentheses give the uncertainties in the last digits. Folding the X-ray data on this ephemeris reveals eclipse times in the X-ray and optical that are within 0.05 in phase. Taking into account the error on the optical ephemeris the eclipse times may well be at exactly at the same phase.

\subsubsection{X-ray and optical spectra}

\begin{figure}
   \centering
   \includegraphics[width=6cm, angle=-90]{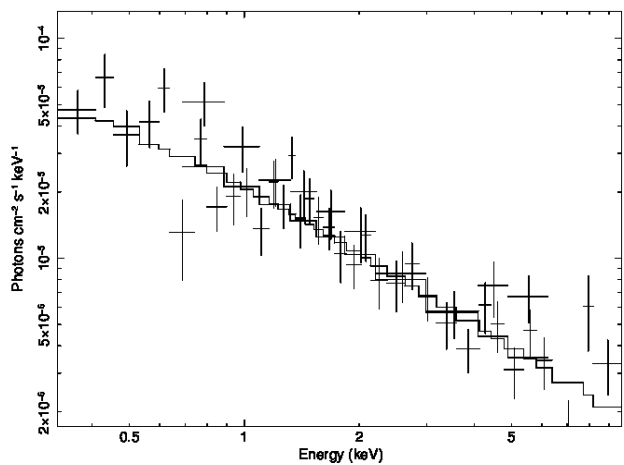}
   \caption{High state EPIC X-ray spectrum of \object{3XMM J183333.1+225136} fitted with a power law.  The fit parameters are found in Table~\ref{tab:SpecFit}.   }
              \label{fig:1833spec}%
    \end{figure}

The best results from the X-ray spectral fitting of \object{3XMM J183333.1+225136} are given in Table~\ref{tab:SpecFit}.  Only the results from the power law and the mekal model are shown as these provided the best fits to the low signal to noise data. The best-fit temperature is high indicating the presence of a partial covering or warm absorber \cite{muka17}. The choice of the specific model used to fit the data does not significantly change the 0.2$-$10 keV flux. Fig.~\ref{fig:1833spec} shows the X-ray spectrum of \object{3XMM J183333.1+225136} from the pulse, corresponding to the pulse maximum ($\phi\sim$0.85-1.3), so as to increase the signal to noise ratio.

\begin{table*}
\caption{Best fits to the CV spectra. All the EPIC data were used for spectral fitting when available, but only MOS 2 data was available for the \object{3XMM J184916.1+652943} spectrum. We report the object name, the interstellar absorption ($\times$ 10$^{22}$ atom cm$^{-2}$), the mekal temperature (or the power law index),  the reduced chi-squared and the number of degrees of freedom. An estimate of the flux in the 0.2--10.0 keV band ($\times$ 10$^{-12}$ erg cm$^{-2}$ s$^{-1}$) is also reported. All the errors are given for 90\% confidence for one interesting parameter.}             
\label{tab:SpecFit}      
\centering          
\begin{tabular}{c c c c c c}     
\hline\hline       
Source & n$_H$ & kT (keV) & $\Gamma$ & $\chi^{\scriptscriptstyle 2}_{\scriptscriptstyle \nu}$ (dof) & Flux \\
\hline
\object{3XMM J183333.1+225136} & 0.05$^{\scriptscriptstyle +0.03}_{\scriptscriptstyle -0.03}$ & $>$41.9 &  & 1.64 (42) & 0.18 \\
  & 0.02$^{\scriptscriptstyle +0.05}_{\scriptscriptstyle -0.02}$ & & 1.08$^{\scriptscriptstyle +0.16}_{\scriptscriptstyle -0.13}$  & 1.54 (42) & 0.19 \\
\hline
\object{3XMM J184916.1+652943}& - & $>$15.0 &  & 0.90 (24) & 2.1 \\
& - & & 1.33$^{\scriptscriptstyle +0.13}_{\scriptscriptstyle -0.13}$ &  0.80 (23) & 2.1 \\
\hline                   
\end{tabular}
\end{table*}

\begin{figure}
\resizebox{\hsize}{!}{\includegraphics[clip=]{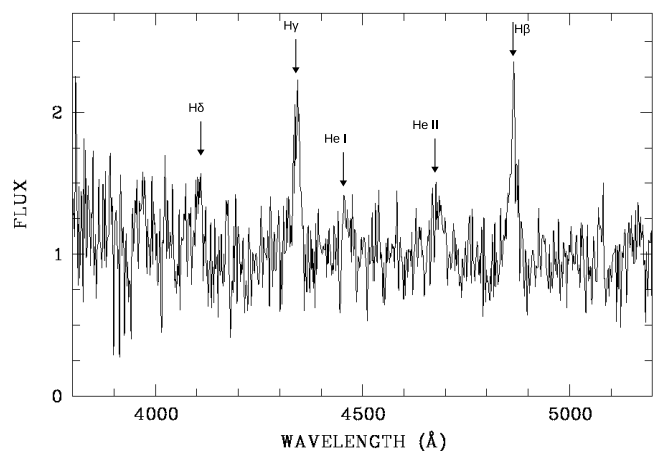}}
\caption{Mean LBT spectrum of \object{3XMM J183333.1+225136} obtained 20-21 November 2016. Flux units are $10^{-17}$\,erg cm$^{-2}$ s$^{-1}$ \AA$^{-1}$. The data of both telescopes were averaged and re-binned to 2\AA. The most prominent lines are indicated (see text for further details).
\label{f:lbt1833}}
\end{figure}

Figure~\ref{f:lbt1833} shows the blue part of the \object{3XMM J183333.1+225136} spectrum.
The continuum magnitude was found to be about $V \simeq 21.5$ (AB magnitude). The object has
an almost flat continuum rising slightly towards the blue. Strong, asymmetric Balmer emission lines (H$\beta$, H$\gamma$, H$\delta$)
and weaker emission lines of neutral (4471\AA) and ionised helium
(4686\AA) are also visible. 

\subsection{3XMM J184916.1+652943}
\label{sec:results1849}

\subsubsection{X-ray and optical variability}

\begin{figure}
   \centering
   \includegraphics[width=10cm, angle=0]{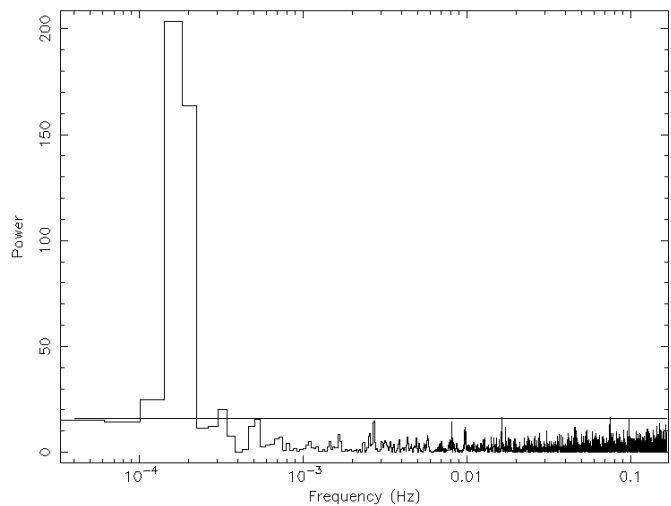}
   \caption{Power spectrum of \object{3XMM J184916.1+652943}. White noise has been subtracted. The horizontal line indicates the 99.95\% significance assuming pure Poisson noise.}
              \label{fig:1849power}%
    \end{figure}

\begin{figure}
   \centering
   \includegraphics[width=9cm, angle=0]{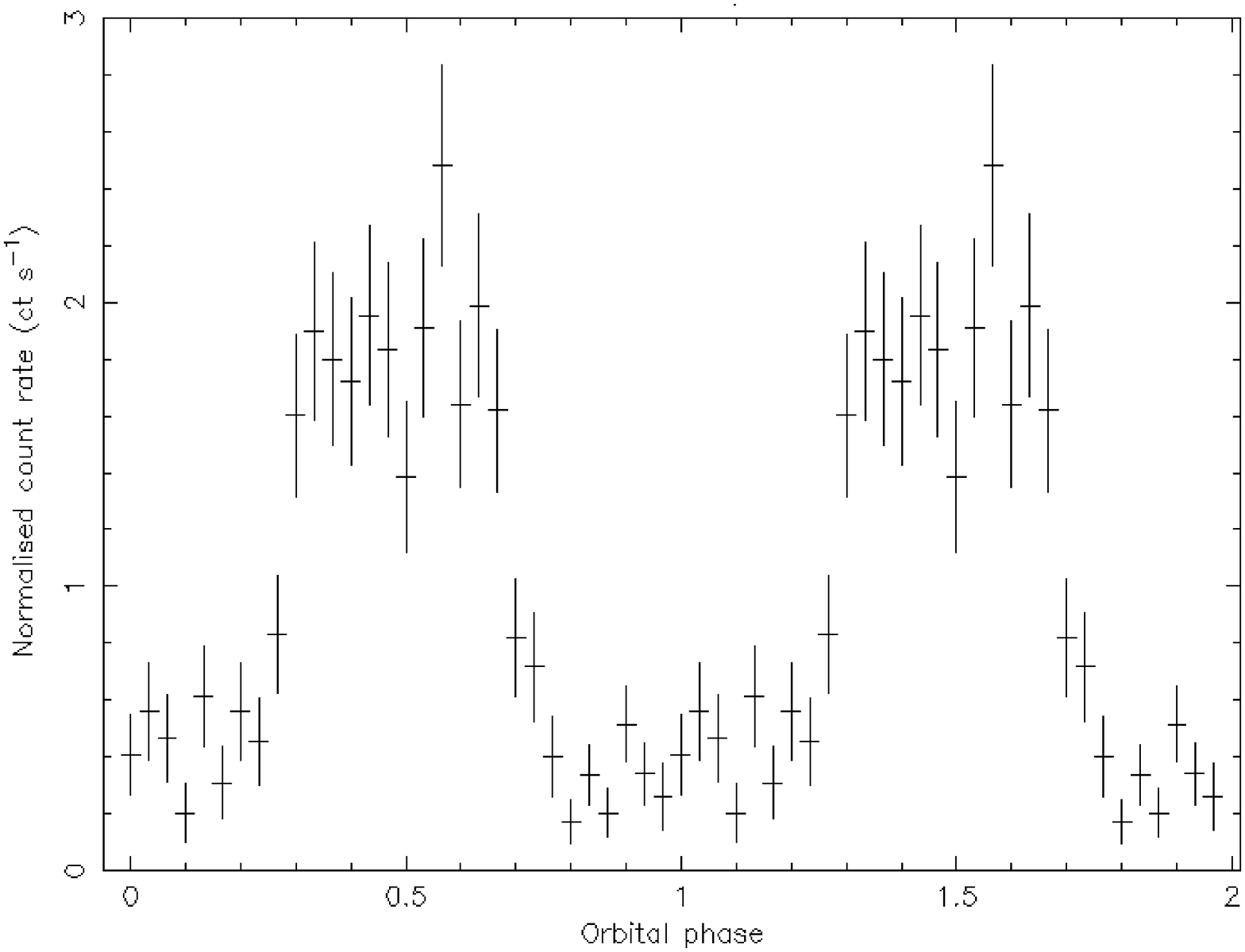}
   \caption{0.2--12.0 keV MOS2 X-ray lightcurve of \object{3XMM J184916.1+652943}, folded on the period of 5800 s and shown with bins of 193.3 s. Two periods are shown for clarity.}
              \label{fig:CV1FoldedLC}%
    \end{figure}

Two significant peaks are found in the power spectrum of \object{3XMM J184916.1+652943} with the strongest peak at P $\simeq$ 5800 $\pm$40 s (3 $\sigma$ error) which may be the orbital period. The third harmonic is at $\sim$1933 s (P/3). The fifth harmonic, $\sim$ 1160 s (P/5), which is also expected for a square shape profile, has a significance less than 3.5 $\sigma$, see Fig.~\ref{fig:1849power}.  A couple of other periods below 100 s touch the 3.5 $\sigma$ significance line, but folding on these periods shows no evidence for periodic modulation. We show the lightcurve folded on the 5800 s period in Fig.~\ref{fig:CV1FoldedLC}.  The system is bright for approximately half of one orbital period.  The X-ray rise lasts approximately 0.05 of the orbit, similar to \object{3XMM J183333.1+225136}, however, the rest of the orbit shows significant emission at 0.058$\pm$0.020 count s$^{-1}$ (99\% confidence).  As for \object{3XMM J183333.1+225136} we compared the X-ray profiles in the bands 0.2--2.0 keV and 2.0--12.0 keV and to within the errors we see no differences. The pulsed fraction for the 0.2--2.0 keV band is 83$\pm$2\% and 86$\pm$2\% for the 2.0--12.0 keV band.

The object PSO\,J184916.434+652943.073, which is at a distance of 2.2\arcsec\ from the X-ray source \object{3XMM J184916.1+652943}, appears to be its optical counterpart due to its blue nature. It has {\em Pan-STARRS} mean aperture AB magnitudes of $19.92\pm0.04, 19.12\pm0.02, 
18.80\pm0.04$ in $g'$, $r'$ and $i'$ respectively. The field around \object{3XMM J184916.1+652943} was also covered by the {\em Catalina Sky
Survey} \citep{drak09}  over 46 epochs between MJD +54951.430757 (2009 April 30) and +56580.159761 ($\sim$2013 October 15). The counterpart was detected and
showed significant variability with the white light magnitudes ranging from 20.0 to 18.3.

Differential aperture photometry of \object{3XMM J184916.1+652943} was performed with respect
to the PanSTARRS object PSO\,J184906.375+652916.478 which has a mean $g$-band aperture 
magnitude of 15.885\,mag. The median photometric uncertainty was about 0.1 mag.
The target did not display significant variability
in the reduced lightcurves. In all data sets the median magnitude difference 
between the target and the comparison star was 3.8 magnitudes indicating that the target had a median magnitude
of $g'=19.7$ during the STELLA observations, similar to the {\em PanSTARRS} photometry.

\begin{figure}
\resizebox{\hsize}{!}{\includegraphics[clip=]{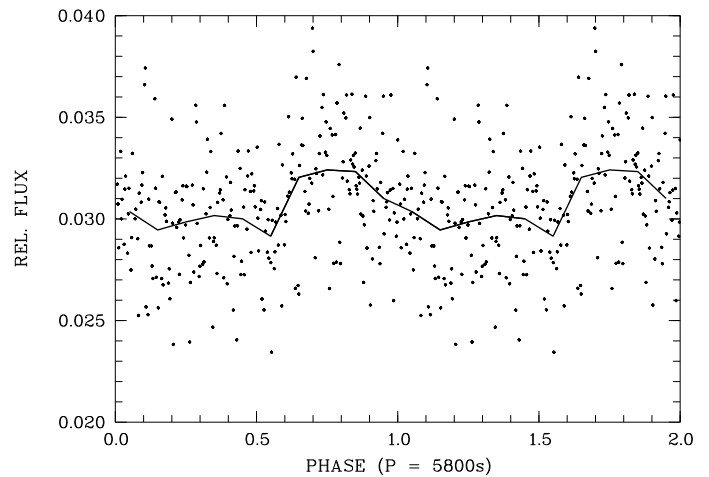}}
\caption{STELLA/WiFSIP photometry of \object{3XMM J184916.1+652943}  obtained in July 2014 folded over
  the X-ray period of 5800\,s. The zero point was chosen arbitrarily. The solid line shows the lightcurve with the data points grouped into ten bins.
\label{f:binlc1849}}
\end{figure}

We folded the optical photometric data on the period determined from the X-ray data (5800\,s), see Section~\ref{sec:results1849} and binned the data into ten phase bins to generate the lightcurve shown in Fig.~\ref{f:binlc1849}.  This data is variable at the 10\% level.  Periodic variability with a pulsed fraction below 5\% can be seen in the lightcurve. The pulse lasts about 50\% of the cycle as in the case of the X-ray lightcurve. 

\subsubsection{The X-ray and optical spectra}

\begin{figure}
   \centering
   \includegraphics[width=6cm, angle=-90]{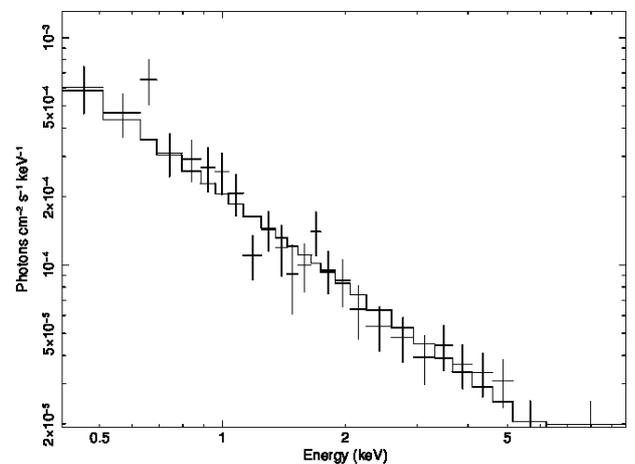}
   \caption{MOS2 X-ray spectrum of \object{3XMM J184916.1+652943} fitted with a power law.  The fit parameters are found in Table~\ref{tab:SpecFit}.   }
              \label{fig:1849spec}%
    \end{figure}

The spectrum of \object{3XMM J184916.1+652943} is as hard as the spectrum of \object{3XMM J183333.1+225136}, as can be seen in Table~\ref{tab:SpecFit} and Fig.~\ref{fig:1849spec}. The X-ray flux is $\sim$2.1 $\times$ 10$^{-12}$ erg cm$^{-2}$ s$^{-1}$ (0.2-10.0 keV band).

The optical spectrum of the optical counterpart to \object{3XMM J184916.1+652943} was found to be much fainter than expected from STELLA photometry. 
The reduced spectrum was very noisy,
displayed a rather blue continuum, likely due to the white dwarf, but did not
reveal any obvious spectral signature that could be used to better identify
and characterise the object. The mean-through slit magnitude was about V=21.5,
hence about 1.5-2 mag fainter than during time-resolved photometry and in the
PanSTARRS catalogue.

\section{Discussion}

\subsection{3XMM J183333.1+225136}
\label{sec:discuss3XMMJ183333.1}

The \object{3XMM J183333.1+225136} X-ray lightcurve is very similar to that of a number of other polars, for example, EP DRa \citep{rams04}, DP Leo \citep{schw02}, WW Hor \citep{pand02}, and V2301 Oph \citep{rams07} amongst others, with a rapid rise to the pulse that lasts approximately half the orbital period, and showing a central dip due to the eclipse of the white dwarf.  In addition, the fact that its X-ray spectrum is also well fitted with a high temperature mekal model supports a polar classification.  High temperature emission is generated in the post-shock flow in the accretion column of magnetic cataclysmic variables. In the case of low quality spectra, this simple model is often appropriate \cite[e.g.][]{home05,webb13,serv08b}. However, whilst the X-ray lightcurve strongly resembles that of polars, the low signal to noise makes it difficult to completely rule out the possibility that the source is an X-ray binary. Black hole X-ray binaries would not show such strong pulsations and if the period proposed to be the orbital period is indeed as short as suggested, this would make the system the shortest period black hole X-ray binary known \citep[e.g.][version 7.2]{ritt03}. The hard X-ray spectrum could indicate a neutron star X-ray binary, but all pulsating neutron star X-ray binaries with such short orbital periods are accreting millisecond pulsars \citep[][version 7.2]{ritt03} and do not show a pulse lasting approximately one hour. It is therefore more likely that \object{3XMM J183333.1+225136} is a cataclysmic variable of the polar type.

If \object{3XMM J183333.1+225136} is a polar, its orbital period (2.15 h) would be at the lower limit of the two- to three-hour period gap observed for CVs \citep[e.g.][]{knig06}, where gravitational radiation is sufficiently important that the separation of the two stars diminishes to bring them back into contact so that accretion can recommence. However, as polars do not seem to experience a large discontinuous change in angular momentum loss, they do not show such a period gap \citep{gaen09}. As we see an eclipse, the inclination is likely to be $>$80$^\circ$.


\cite{knig11} give an empirical relation between the CV orbital period and the nature of the secondary star. Using their relation, we found that the secondary mass and radius are M$_2$=0.17 M$_\odot$ and R$_2$=0.21 R$_\odot$ respectively.  With regards to the white dwarf (primary), it is difficult to determine an exact mass from the data in hand. The primary mass should be weakly dependent on the orbital period. It is 0.75$\pm$0.05 M$_\odot$  for CV primaries with orbital periods between approximately one and six hours \citep{knig11}.


With regards to the distance to this object, given the Galactic longitude and latitude (Galactic coordinates, $l$=51.745$^\circ$ $b$=+13.950$^\circ$), it is not towards the Galactic centre. At large distances the object would be high out of the Galactic plane and in the Galactic halo. It is unlikely that a CV is found in the Galactic halo and therefore must be fairly local ($<$1.2 kpc).  If it is within this distance, then the X-ray luminosity is $\lesssim$3 $\times$ 10$^{31}$ erg s$^{-1}$, which is similar to other polars, see references above. \cite{warn87} determined an empirical relationship between the orbital period and the absolute V-band magnitude of the system. Using the relationship for dwarf-novae in a low state, which approximates to the relationship for polars \citep{warn87}, the absolute magnitude is M$_V\sim$9. If the system is situated at $\sim$1 kpc, this would give a m$_V\sim$19, which means that it would have been detected in the V-band observations made in the OM, see Table~\ref{tab:OMdata}. The object should then be situated $>$2.5 kpc away for us not to have detected it, if it was observed in the low state. The strong emission lines in the optical spectrum confirms that our optical source is the optical counterpart of the X-ray source.  Further, the magnetic CVs that do not possess an accretion disc, cannot be plotted on the magnitude-period diagrams even when their distances are known, but using estimates of their mass transfer, however, they tend to fall among the dwarf novae at minimum light \citep{warn87}. This magnitude is therefore only an estimate and does not call the distance into question. \cite{knig06} provides empirical relations of the absolute infra-red magnitude (J, H and K bands) versus period, which would give another method to estimate the distance, but no archival data for \object{3XMM J183333.1+225136} exists in these bands.  The very blue nature of this CV (see Table~\ref{tab:OMdata}), the optical phase-resolved variability, the occurrance of high and low states, and the emission lines in the optical spectrum all support the polar identification.

\subsection{3XMM J184916.1+652943}

The lightcurve and spectrum observed for this source support the magnetic cataclysmic hypothesis. As no rotation period at high frequencies was identified in the power spectrum, it is likely that this source is a polar.  Its short period of 1.6 hours (97 minutes) places it close to the period minimum for CVs, that is, observed around 80 minutes or theoretically predicted to be around 65 minutes \citep{knig11}.  The absence of an eclipse constrains the system to have an inclination $\lesssim$75$^\circ$.  As for \object{3XMM J183333.1+225136} it is difficult to completely rule out the X-ray binary hypothesis, but for the same arguments as in Section~\ref{sec:discuss3XMMJ183333.1}, it is more likely that \object{3XMM J184916.1+652943} is a polar.

The empirical relation in \cite{knig11} indicates that the mass and the radius of the secondary star are around M$_2$=0.12 M$_\odot$ and  R$_2$=0.16 R$_\odot$.  Given that this object also has high Galactic latitude (Galactic coordinates, $l$=95.761$^\circ$ $b$=+24.734$^\circ$), it is not towards the Galactic centre and must be closer than 1 kpc to be in the disc and not in the Galactic halo. Again, no infra-red data is available to refine this distance estimate. This object is therefore likely to be local with an X-ray luminosity of $\lesssim$2.8 $\times$ 10$^{32}$ erg s$^{-1}$, a reasonable value for a polar. The faint phase observed from \object{3XMM J184916.1+652943}  then has an unabsorbed flux of 8.2$\times$ 10$^{-14}$ erg cm$^{-2}$ s$^{-1}$ (0.2--10.0 keV) and an X-ray luminosity of 1$\times$ 10$^{31}$ erg s$^{-1}$. This is too high to be coming from the secondary star, if it is for example a late-type active star. However, a second pole coming into visibility, for example, like BL Hyi \citep{beue89} could then explain this level of emission. 

We note that \object{3XMM J184916.1+652943} is probably the {\em RASS} ({\em ROSAT all-sky survey}) source 1RXS J184915.9+652947 that was studied by
\cite{Fuhr03}, as the two X-ray fluxes are very similar (to within a factor of two). The {\em ROSAT} and the {\em XMM-Newton} positions agree to within their errors.  \cite{Fuhr03}  identified their flare-like variable object with the bright star HD 175199 (B=8.9, V=8.6). This star is found at a distance of 85\arcsec\ and therefore the association appears unlikely. Interestingly, this X-ray source is not listed in 2RXS, the recently published catalogue based on a re-processing of the {\em RASS} by \cite{boll16}.

\subsection{Population of magnetic CVs}

These two new short orbital period magnetic CVs both have very hard spectra. Polars were initially identified due to their soft X-ray spectra, detected with {\em Rosat} (0.1--2.4 keV), for example, \cite{buck93,sing95,maso95}, but more recently, hard X-ray satellites such as {\em Integral}, {\em Swift BAT} and {\em NuSTAR} have been identifying new polars, for example \cite{barl06,hong16}.  There are 148 polars listed in the Catalogue of Cataclysmic Binaries \citep[e.g.][version 7.2]{ritt03} and 11 polars have now been selected due to their hard X-ray spectrum \citep{bern17}.  This has been proposed to be due to the fact that the Galactic centre region, which is often surveyed by these hard X-ray satellites, may harbour a higher fraction of magnetic CVs with higher mass white dwarfs than elsewhere in the Galaxy, which in turn produce harder X-rays \citep{hong16}.  However, the magnetic CVs identified here are not in the Galactic centre.  Nonetheless, other magnetic CVs without soft X-ray emission have been found in the {\em XMM} catalogue \citep[e.g.][]{voge08,rams09}, which may indicate that such hard magnetic CVs are not so rare.

These two new short period magnetic CVs were identified through studying source products from a very small fraction of the 3XMM catalogue. This fact indicates that such a catalogue is likely to contain a large number of similar objects that are only slightly fainter than the detection limit of the {\em Rosat} all sky survey \citep[][]{voge99} and slightly fainter than the detection limits of X-ray telescopes sensitive to harder X-rays such as the {\em Swift BAT} or {\em INTEGRAL}.  Given that the 3XMM catalogue covers only a few percent of the sky, future sensitive all sky surveys such as the {\em eROSITA} project \citep{pred13}  should be very successful at uncovering large numbers of such sources, see also \cite{schw12}, where both of these sources are bright enough to be detected by {\em eROSITA}.

\begin{acknowledgements}
 Firstly we would like to thank the anonymous referee for their useful feedback on this paper.  This research has made use of data obtained from the 3XMM {\em XMM-Newton serendipitous source catalogue} compiled by the ten institutes of the {\em XMM-Newton Survey Science Centre} selected by ESA.  Based partly on data obtained with the STELLA robotic telescopes in Tenerife, an AIP facility jointly operated by AIP and IAC.  Based on observations obtained with {\em XMM-Newton}, an ESA science mission with
instruments and contributions directly funded by ESA Member States and NASA.  We thank E.~Farina and J.~Storm for obtaining the spectra at the LBT.  The LBT is an international collaboration among institutions in the United
States, Italy and Germany. LBT Corporation partners are: The University of
Arizona on behalf of the Arizona university system; Istituto Nazionale di
Astrofisica, Italy; LBT Beteiligungsgesellschaft, Germany, representing the
Max-Planck Society, the Astrophysical Institute Potsdam, and Heidelberg
University; The Ohio State University, and The Research Corporation, on behalf
of The University of Notre Dame, University of Minnesota, and University of
Virginia.

\end{acknowledgements}


\bibliographystyle{bibtex/aa} 

\bibliography{30974_corr.bib}




\end{document}